\documentstyle[aps,floats,epsf,color]{revtex}
\begin{document}
\baselineskip=12pt
\def\be{\begin{equation}}
\def\ee{\end{equation}}
\def\ba{\begin{eqnarray}}
\def\ea{\end{eqnarray}}
\def\peffec{\peight{\bearst\eearst}\hfill\peleven}
\def\pspace{\peight{\bearst\eearst}\hfill}
\def\ptwelve{\parbox{12cm}}
\def\peight{\parbox{8mm}}
\twocolumn[\hsize\textwidth\columnwidth\hsize\csname@twocolumnfalse\endcsname

\title{\large \bf Lensing effects in inhomogeneous cosmological models}
\author{Sima Ghassemi$^{1}$\thanks{E-mail: ghassemi@ipm.ir},
Salomeh Khoeini Moghaddam$^{2}$\thanks{E-mail: skhoeini@tmu.ac.ir
}, and Reza Mansouri$^{1,3}$\thanks{E-mail:   mansouri@ipm.ir}}
\address{$^{1}$School of Astronomy and Astrophysics, Institute for
Research in Fundamental Sciences (IPM),\\ P. O. Box 19395-5531,
Tehran, Iran,}
\address{$^{2}$Physics Dept., Faculty of Science, Tarbiat Mo'alem
University, Tehran, Iran,}
\address{$^{3}$Department of Physics, Sharif University of Technology,\\
Tehran 11365-9161, Iran}

\vskip 1cm

\maketitle
\begin{abstract}

Concepts developed in the gravitational lensing techniques such as
shear, convergence, tangential and radial arcs maybe used to see
how tenable inhomogeneous models proposed to explain the
acceleration of the universe models are. We study the widely
discussed LTB cosmological models. It turns out that for the
observer sitting at origin of a global LTB solution the shear
vanishes as in the FRW models, while the value of convergence is
different which may lead to observable cosmological effects. We
also consider Swiss-cheese models proposed recently based on LTB
with an observer sitting in the FRW part. It turns out that they
have different behavior as far as the formation of radial and
tangential arcs are concerned.\\

\end{abstract}
\newpage
]

\section{Introduction}

There is an on-going debate in the community wether
inhomogeneities on scales up to several hundred mega parsecs could
account, at least partially, for the observed dimming of the SNe
Ia [Celerier\cite{celerier}, Sarkar \cite{sarkar}]. Irrespective
of successes of these models to explain the dark energy, it is
desirable to have some independent cosmological tests of these
models. The obvious effect of the inhomogeneities is the light
propagation which may differ significantly in a clumpy universe
compared with a homogeneous FRW model. Recently Vanderveld et al
\cite{vander2}, being interested in the possibility of explaining
the supernova data, have studied light propagation in one of the
Swiss cheese models recently proposed \cite{Kolb} using weak field
gravitational lensing techniques. We are interested in the
potential gravitational lensing effects in inhomogeneous models to
see if there are crucial effects which could constrain or
rule out some of the models. \\
In this paper we focus on LTB-based inhomogeneous models. The LTB
solution of Einstein equations describes an inhomogeneous
spherically symmetric dust filled cosmological model with a
distinguished center of symmetry. Having a center of symmetry,
thus not reflecting the homogeneity of the universe at large,
these models are mainly used as toy models to study the local
inhomogeneities of the universe. After a short review of LTB
inhomogeneous solutions, and Swiss-cheese models based on it, we
look at the general question of how different does a LTB model
behave relative to a FRW model as far as gravitational lensing
concepts are concerned (section \ref{global}). For an observer at
the origin of a flat LTB metric, the shear vanishes as in the case
of a FRW metric. However, the convergence is different from that
of the FRW due to the r-dependence of metric coefficients and the
corresponding scale factor. The vanishing of the shear is only
true for an observer at the origin of LTB. The more general and
interesting case of an off-center observer has to be treated
separately. This case, however, maybe treated in Swiss cheese
models based on LTB which is the subject of the section
\ref{Swisscheese}, where arcs in strong lensing regime are
investigated for two different Swiss-cheese models, assuming the
observer and the source are sitting in the cheese (FRW metric).
While the onion model \cite{mansouri nottari} turns out to produce
neither tangential nor radial arc, the model proposed by Marra,
Kolb, Matarrese and Riotto \cite{Kolb} do produce both radial and
tangential arcs. The models we have considered could be looked at
as providing different density profiles of a lens, to be compared
with the profiles published so far and mainly based on
simulations. The Swiss cheese models, e.g., provide us with
density profiles which maybe compared for special parameters to
the NFW one which is a phenomenological density profile well
supported by the simulations\cite{bartelmann96}.

\section{A review on inhomogeneous LTB models}\label{review}

The LTB metric ($c=1$) in co-moving and proper time coordinates with zero
cosmological constant is given by
\begin{equation}\label{general form}
ds^2=-dt^2+S^2(r,t)dr^2+R^2(r,t)(d\theta^2+\sin^2\theta d\phi^2).
\end{equation}
Considering dust stress-energy tensor, Einstein's equations imply the
following constraints:
\begin{eqnarray}\label{metric coeficients}
S^2(r,t)&=&\frac{R^{'2}(r,t)}{1+2E(r)},\\
\frac{1}{2}\dot{R}(r,t)&-&\frac{GM(r)}{R(r,t)}=E(r),\\
4\pi\rho(r,t)&=&\frac{M^{'}(r)}{R^{'}(r,t)R^2(r,t)},
\end{eqnarray}
where dot and prime denote partial derivatives with respect to t
and r respectively.  Function $\rho(r,t)$ is energy density of the
matter. Functions $E(r)$ and $M(r)$ are left arbitrary. $M(r)$ can
be interpreted as the mass inside of co-moving sphere with
coordinate radius $r$. Assuming $M^{'}(r)>0$, $M(r)$ can be chosen
as
\begin{equation}
M(r)=\frac{4\pi}{3}M_0^4r^3.
\end{equation}
Now, taking this as the definition of the coordinate radius, we
may write solutions of Einstein's equations depending on $E(r)$ in
the following ways:

1. For $E>0$ the solution is
\begin{eqnarray}\label{E>0 1}
  R &=& \frac{GM(r)}{2E(r)}(\cosh u-1) \\
  t-t_n(r) &=& \frac{GM(r)}{[2E(r)]^{3/2}}(\sinh u-u);
\end{eqnarray}

2. For $E=0$ we have
\begin{equation}\label{E=0}
    R(r,t)=[\frac{9}{2}GM(r)]^{1/3}[t-t_n(r)]^{2/3};
\end{equation}

3. Finally for $E<0$ the solution is
\begin{eqnarray}\label{E<0 1}
  R &=& \frac{GM(r)}{-2E(r)}(1-\cos u) \\
  t-t_n(r) &=& \frac{GM(r)}{[-2E(r)]^{3/2}}(u-\sin u).
\end{eqnarray}

The so-called bang time function $t_n(r)$ is an integration
constant, indicating different singularities defined by $t =
t_n$\cite{Khosravi}.

 As we are going to discuss phenomenon
far from these singularities, we may assume $t\gg t_n$. Therefore, we will assume from now on $t_n = 0$.

\subsection{Small-u approximation}\label{perturb}

Relation between the coordinate $r$ and the parameter $u$ in the
non-flat LTB cases is not trivial. It has been shown that for the
special case of $E(r)$ being a trigonometric function of r, small
u approximation is even valid for enough large r \cite{mansouri
nottari}. This, however, is not in general the case as may be seen
for a polynomial function $E(r)$. However, to simplify the
calculation, we are going to assume $u$ to be a small parameter.
This approximation, which can describe the dynamics even when
$\delta\rho/\rho\gg1$, allows to solve the Einstein's equations
\cite{mansouri nottari,notari}.

 Physically $u^2$ is related to spatial curvature
$E(r)/r^2$ \cite{mansouri nottari,notari}. In addition we assume
$E>0$, leading to the following equations:
\begin{eqnarray}
  R(r,t) &=& \frac{2\pi r}{3k(r)}(\cosh u-1)\label{E>0 21}, \\
  \tilde{M}t &=& \frac{\sqrt{2}\pi}{3k(r)^{3/2}}(\sinh u-u)\label{E>0 22},
\end{eqnarray}
where
\begin{eqnarray}
k(r)&:=&\frac{E(r)}{\tilde{M}^2r^2}\:, \\
\tilde{M}&:=&\frac{M_0^2}{m_{pl}}\:,\hspace{1.5cm}(m_{pl}=\sqrt{\frac{1}{G}}).
\end{eqnarray}
Keeping next to leading terms in u, we obtain from (\ref{E>0 21})
and (\ref{E>0 22})
\begin{eqnarray}
R&\approx&\frac{\pi r}{3k(r)}u^2\left(1+\frac{u^2}{12}\right),\nonumber\\
\tau^3&:=&\tilde{M}t\approx\frac{\pi\sqrt{2}}{18k(r)^{3/2}}u^3\nonumber.
\end{eqnarray}
As mentioned before, the relation between parameter $u$ and spatial
curvature $k(r)$ is given by
\begin{eqnarray}
u=\frac{18}{\pi\sqrt{2}}\tau\sqrt{k(r)}.\nonumber
\end{eqnarray}
Therefore, the small u approximation is valid when

\begin{equation}
u=\gamma\tau\sqrt{k(r)}\ll1,
\end{equation}

where

\begin{eqnarray}
  R_2:=\frac{1}{12}=0.08,\hspace{1.5cm}
  \gamma:=\left(\frac{9\sqrt{2}}{\pi}\right)^{1/3}\approx1.59.\nonumber
\end{eqnarray}

Substituting u yields,
\begin{eqnarray}
R(r,t)=\frac{\pi}{3}\gamma^2\tau^2r[1+R_2\gamma^2\tau^2k(r)].
\end{eqnarray}

\subsection{Swiss-cheese model}\label{swiss}

The inhomogeneous metrics may be used to model universe in
different ways. The direct way is to take an inhomogeneous metric,
say a LTB solution of Einstein equation, as the model universe and
see the effect of lensing in it. One may, however, devise a
so-called Swiss-cheese model in which the bulk (cheese) is
represented by a matter-dominated flat homogeneous FRW model and
the spherically symmetric holes are constructed using a specific
LTB solution. The holes which represent the inhomogeneities are
distributed randomly in the bulk, so the model is isotropic and
homogenous on average. The matching of the inhomogeneous holes to
the FRW bulk must be handled with care \cite{khak}. Depending on
different types of LTB
solutions, one may construct different Swiss-cheese models.\\
Biswas et. al \cite{notari} study a Swiss-cheese model in which
the holes are represented by a LTB metric in the small u
approximation regime (section \ref{perturb}). They choose $M_0$ in
such a way that the coordinate density, $M_0^4$, coincides with
the average density ($\rho_0$) at present time $t_0$:
\begin{eqnarray}\label{norm}
M_0^4=\rho_0=\frac{M_p^2}{6\pi t_0^2},
\end{eqnarray}

or

\ba \tau_0^2=t_0\tilde{M}=\frac{1}{\sqrt{6\pi}}.\nonumber\ea

The matching conditions imply \cite{vander}
 \ba k'(L)=0,\nonumber\ea
where L is the comoving radius of the hole and prime means derivative
with respect to $r$. Using the above normalization (\ref{norm})
for $\Omega_k\ll 1$ ($\Omega_k$ is curvature abundance of the homogeneous
universe at the same time)we arrive at \ba k(L)=\frac{4\pi}{3}\Omega_k. \nonumber\ea In
order to be consistent with CMB we choose \ba k(L)=0.\nonumber\ea
According to \cite{vander}, continuity at the origin implies
another constraint on curvature:
 \ba k'(0)=0\nonumber\ea

Recently Marra et.al \cite{Kolb} defined a different Swiss-cheese
model in which arbitrary number of spherical holes with different
size and density profile are distributed in the cheese. The cheese
evolves as FRW while the holes evolve differently. At the boundary
of the holes, as a consequence of the boundary conditions, the
average mass density, defined by
$\overline{\rho}=\frac{3}{R(r,t)^3}\int_0^r\rho(r,t)R^2R'dr$,
coincides to the FRW density, and E(r) has to go to zero. As far
as local physics is concerned, the hole has no effect on the
observer outside it.

\section{LTB universe - Observer at the origin}\label{global}

Let us first study the lensing effect due to the global inhomogeneity in a
flat LTB solution($E(r)=0$):

 \be \label{LTBmetric} ds^2=dt^2-R'(t,r)^2dr^2-R(t,r)^2d\Omega^2.\ee
We do not assume any single or multiple lens but are interested in the global
effect of bending of light rays due to the LTB inhomogeneities. For simplicity, we
place the observer at the event O (center of the inhomogeneous region) with 4-velocity
$u^{\alpha}_o$, $u^{\alpha}_ou_{\alpha o}=1$. Choosing the affine parameter of the
rays, $\lambda$, at O such that (1) $\lambda = 0$ at origin,
 (2) $\lambda$ increases to the past and (3)
 $k_{\alpha}u^{\alpha}_o\mid_O = -1$, then
 $k^{\alpha} = \frac{dx^{\alpha}}{d\lambda}$ is past directed. Using the dimensionless
$k^{\alpha}$, the corresponding wave vector is then defined by $-\frac{\omega_o}{c}k^{\alpha}$,
where $\omega_o$ is the frequency of the wave measured by the observer at O.\\
Let $\gamma_0$ be a ray and $u^{\alpha}$ on $\gamma_0$ be the
result of the parallel propagated four velocity of the observer,
$u^{\alpha}_o$. The orthonormal bases along $\gamma_0$ on the lens
plane are $E^{\alpha}_1$ and $E^{\alpha}_2$.  The deviation
vectors of the beam centered on $\gamma_0$ can then be written as
$Y^{\alpha}=-\xi_1 E^{\alpha}_1-\xi_2 E^{\alpha}_2-\xi_0
k^{\alpha}$, where $\xi_1$ and $\xi_2$ are called the screen
components of the corresponding separation vector of two
neighboring light rays\cite{Seitz}. For the above metric these
vectors are derived as \be\label{orthonormal bases-LTB}
 E_1^{\alpha}=\left[0,0,\frac 1
{\sqrt{g_{\theta\theta}(z)}},0\right],\:\:\:
E_2^{\alpha}=\left[0,0,0,\frac 1 {\sqrt{g_{\phi\phi}(z)}}\right],\ee
and \be
k^{\alpha}=(1+z)\left[-1,\frac{1}{\sqrt{g_{rr}(z)}},0,0\right], \ee
where $z$ is the red-shift of the source (O) defined as:
 \be\label{red-shift}
(1+z)=\frac{(k_{\alpha}u^{\alpha})_e}{(k_{\alpha}u^{\alpha})_o}.\ee

The evolution of $\mathbf{\xi}=(\xi_1,\xi_2)$ is given by the
following equation of geodesic deviation\cite{gravitational lenses}:

\be
\label{xi-evolution}\ddot{\mathbf{\xi}}(\lambda)=\mathcal{T(\lambda)}\mathbf{\xi}(\lambda),
\ee where $\mathcal {T}$ is the optical tidal matrix describing
the influence of space-time curvature on the propagation of light:
\be \label{optical tidal metric}\mathcal{T(\lambda)} =\left(
\begin{array}{cc}
\mathcal{R}(\lambda)+ \textit{Re}\mathcal{F}(\lambda)&\textit{Im}\mathcal{F}(\lambda)\\
 \textit{Im}\mathcal{F}(\lambda)& \mathcal{R}(\lambda)-
 \textit{Re}\mathcal{F}(\lambda)\\
\end{array}\right).
  \ee
$\mathcal{R}$ is the socalled source of convergence:

\be\label{convergence definition} {\mathcal{R}}=-\frac 1 2
R_{\mu\nu}k^{\mu}k^{\nu}, \ee where $R_{\mu\nu}$ is the Ricci
tensor of the metric. $\mathcal{F}$ is the source of shear:

\be\label{shear definition} {\mathcal{F}}=-\frac 1 2
C_{\alpha\beta\gamma\delta}\epsilon^{*\alpha}k^{\beta}\epsilon^{*\gamma}k^{\delta},
\ee where $\epsilon^{\alpha}:= E_1^{\alpha}+iE_2^{\alpha}$ and
$C_{\alpha\beta\gamma\delta}$ is the Weyl curvature tensor of the
metric. As expected, for isotropic metrics, like LTB with the
observer at the origin, the source of shear is vanishing:

\be \label{LTB-shear}{\mathcal{F}}_{LTB}=0.\ee

In the case of flat LTB spacetime with corresponding solutions (\ref{E=0}) the relevant source of
convergence is, however, non-vanishing:

\be\label{LTBconvergence}
{\mathcal{R}}=(1+z)^2\left[\frac{\ddot{R}}{R}-\frac{\dot{R}\dot{R}'}{RR'}\right].
\ee

The non-vanishing convergence means that the light coming from a
source at $(r,t)$ is converged by (\ref{LTBconvergence}) when
observed at the origin $O$. To derive the relevant red-shift ($z$)
for each event ($t$ and $r$) we used the numerical code of
(\cite{Khosravi}). We have then plotted the convergence as a
function of increasing $z$ (Fig.1). As it is shown in the picture,
convergence is increasing with z. This convergence is different
from that of the FRW spacetime. To understand this difference one
needs to study different observables such as time delay of images
of a source, which goes beyond the scope of this paper, and we
will go into its detail in a future publication.

\begin{figure}
\epsfxsize=9.0truecm\epsfbox{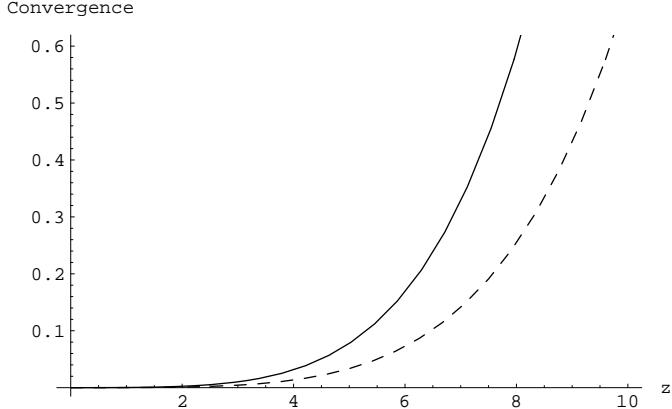} \narrowtext \caption{
Source of convergence as a function of red-shift .The solid curve
stands for the FRW model and the dashed curve for flat LTB.}
\end{figure}

\section{Swiss-cheese model of the universe: Observer in the cheese }\label{Swisscheese}

In the previous section we studied a special problem where the
observer was at the center of inhomogeneities. To be more
realistic we want to study a more general problem where the
observer is placed somewhere between inhomogeneities in the
universe and observes a distant source. The light coming from the
source is passing the inhomogeneous regions in between and reaches
the observer. For simplicity, we consider two different
Swiss-cheese models (section \ref{swiss})) in which the observer
and the source are in the cheese and the light passes through one
of the holes:  Onion model of Biswas et al. \cite{notari} who
derive a perturbative LTB solution of Einstein equations and study
the evolution of the density contrast within the holes, and the
model of Marra et al. \cite{Kolb} who construct a non-perturbative
solution in the holes. In this model the universe is completely
filled with these holes which form a sort of lattice, taking care
of the matching conditions between LTB and FRW metrics on the
boundary of the hole. The hole is almost empty except at the
boundary where the matter is concentrated and
has an average density matching that of FRW density.\\
We will trace light rays within these two Swiss cheese models
coming from a source in the Homogenous Friedman background, the
cheese, passing a lens, a hole, and finally detected by an
observer in the cheese. Deriving the source of convergence and
shear in this case, is not a simple job due to the difficulty of
tracing light rays in the Swiss cheese which is not as
straightforward as in the case of the LTB model with the observer
at the origin.  For arbitrary observer we expect a non-vanishing
shear. The effect of the shear is best studied through observables
such as radial and tangential arcs  \cite{schneider99} of the
distant galaxies. We will first elaborate on some of the basic
definitions and then go on to calculate the arcs in models just
described.

\subsection{Some basic definitions in lensing}

The surface mass density of the lens is defined as

\be \Sigma({\bf{\xi)}})=\int^{+\infty}_{\infty}\rho(r,t) dz,\ee
where $r=\sqrt{z^2+\xi^2}$, $z$ is the coordinate aligned with the
line of sight and $\bf{\xi}$ are the coordinates in the lens plane
(with the origin on the lens). Convergence of the light bundles
made by the lens is defined as follows:

\be\kappa(\xi)=\frac{\Sigma(\xi)}{\Sigma_{critical}}\ee where
$\Sigma_{critical}$ is defined as:
$\Sigma_{critical}:=\frac{c^2}{4\pi G}\frac{D_s}{D_dD_{ds}}$
\cite{schneider99}. Now consider that we are observing a far
galaxy and the light coming from that galaxy to us is bent due to
the lensing of the inhomogeneity which exists in their path.
There are some places in the lens plane where the magnification of
the lens goes to infinity. Then what we see in the lens plane are
some radial or tangential arcs. The magnification of a lens is
defined as follows \cite{schneider99}:

\be \mu\propto \frac 1 {det(\mathcal{A})},\ee where $\mathcal{A}$
is the Jacobian of the transformation of the source plane
coordinates to the lens plane ones. When the determinants of
$\mathcal{A}$ goes to zero we will have the arcs in the lens plane
(called critical curves). Therefore, to derive possible arcs, one
should calculate the eigenvalues of $\mathcal{A}$ first and see if
it can be zero somewhere on the lens plane or not.

The mass inside radius $x$ is described by the dimensionless
function:

\be m(x):=2\int^x_0 \kappa(\xi)d\xi.\ee

It can be easily shown that the eigenvalues  of the $\mathcal{A}$
matrix be derived using this dimensionless function
\cite{schneider99,bartelmann96}:

\be \lambda_r=1-\frac{d}{dx}\frac{m(x)}{x},\:\:
\lambda_t=1-\frac{m(x)}{x^2}.\ee

$\lambda_r$ and $\lambda_t$ are called radial and tangential
critical curves, respectively. When $\lambda_r$  goes to zero the
arcs of an extended source in the lens plane are radial and  when
$\lambda_t$ goes to zero the arcs of an extended source in the lens
plane are tangential.

\subsection{Radial and tangential arcs I: Onion Swiss cheese model}

To study the arcs, we trace light rays coming from a far source (in
the Homogenous Friedman background - the cheese) passing a lens with
perturbative LTB metric, as described in section(\ref{perturb}), and
finally detected by an observer in the cheese. Once we have the
density profile of such a lens, we can immediately derive the place
of arcs and see if there exists both radial and tangential arcs or
not \cite{bartelmann96}. Different density profiles have been
studied before. For example the NFW profile \cite{NFWa} can produce
radial and tangential arcs \cite{bartelmann96}.

Density profile of a LTB lens with perturbative metric (section
(\ref{perturb})) is given by the following relation, derived from
the Einstein equations in the small $u$ approximation:

\be\label{LTBdensityprofile} \rho(r,t)=\frac{M_0^4}{\pi
G(\tilde{M}t)^2\left[1+R_2\gamma^2(\tilde{M}t)^{2/3}A(r)\right]},
\ee where
$\displaystyle{A(r):=\left(\frac{E(r)}{r\tilde{M}^2}\right)'}$.

It seems that the density is decreasing with time, while we expect
the structure grows as a function of time. But the important point
here is that the relevant quantity to study the structure growth is
the density contrast.

The background spacetime is a FRW spacetime. Hence, one can define
the density contrast as the deviation  of the holes density
($\rho(r,t)$) from the cheese density ($<\rho>(t)$, averaged over
the holes)\cite{mansouri nottari}:

\be\label{densitycontrast}
\delta:=\:\frac{\rho(r,t)\:-<\rho>(t)}{<\rho>(t)},\ee defining
$\epsilon(r,t):= R_2\gamma^2(\tilde{M}t)^{2/3}A(r)$, one can
derive the density contrast:

\be\label {LTB densitycontrast}
\delta=\frac{-\epsilon(r,t)}{1+\epsilon(r,t)}.\ee

As Biswas et al. show \cite{mansouri nottari}, $A(r)$ has to be bounded to
get underdensity (voids) and overdensity (structure) regions:

\be\label{densitycontrastbound}
A_{min}<A(r)<A_{max}\:\:\rightarrow\:\:\delta_{min}<\delta<\delta_{max}\ee

As $\epsilon(r,t)$ is an increasing function of time, its sign is crucial
to get the underdensity  and overdensity regions. A negative $\epsilon(r,t)$ gives
a positive density contrast which grows with time (growing structures) and
a positive $\epsilon(r,t)$ gives a negative density contrast growing with
time (growing voids). Therefore, $A_{min}$ should be negative, corresponding to an
overdense region. \\

As it is shown in \cite{mansouri nottari}, the following $E(r)$ can satisfy
the above condition (\ref{densitycontrastbound}):

\be\label{sinuoidal} E(r)=\tilde{M}^2
\frac{A_1L}{2\pi}\:r\sin^2\left(\frac{\pi r}{L}\right),\ee
where $L$ is a typical length of the large scale structures and $A_1$ is
the amplitude of the density oscillations. This may be seen by looking at $R(r,t)$:

\be\label{oscillationsamplitude} R(r,t)\approx
(6\pi)^{1/3}(\tilde{M}t)^{2/3} r \left[1+\frac{A(t)}{2\pi}\frac L
r \sin^2\left(\frac{\pi r}{L}\right)\right],\ee where $A(t):= R_2
\gamma^2 A_1\left(\tilde{M}t\right)^{2/3}$. The density profile
can also be written as
 \be\label{small u LTB density}
\rho(r,t)=\frac{M_0^4}{6\pi
(\tilde{M}t)^2\left[1+A(t)\sin(\frac{2\pi r}{L})\right]}, \ee
showing the significance of $A(t)$ as the amplitude of
oscillations.

The special characteristic of this Swiss cheese model is that each
hole has an onion like density profile which at the large $r$'s goes to a
homogeneous background density.

Now, consider one of these LTB regions as a lens. The
plane perpendicular to the line of sight of the source (which can
be a far galaxy) is called the lens plane. The light passes through
the FRW region and, close to the lens, it is bent due to the density profile of
the LTB hole using the small $u$ approximation, just as in the Schwarzschild case.

In realistic cases one may forget about the time evolution of the lens during the
passage of light. This means the characteristic time of lens evolution is much
greater than the light passage time. In the following we will apply this approximation.


\begin{figure}
\epsfxsize=7.0truecm\epsfbox{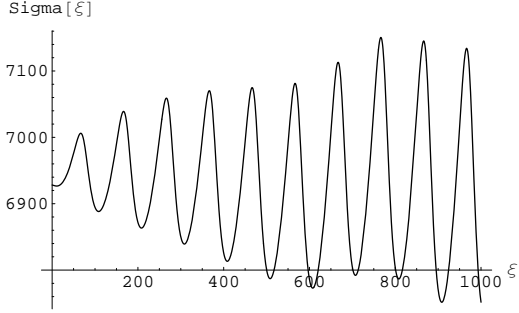} \narrowtext \caption{
Surface mass density of the lens as a function of the coordinates
on the lens plane.}
\end{figure}

We have now all the quantities to calculate $\lambda_r$ and
$\lambda_t$ for the density profile (\ref{small u LTB density}).
Figures 3-6 show the behavior of $m(x)$, $\lambda_r$, and $\lambda_t$
as a function of the distance to the center of the lens in the lens plane, $x$.\\


\begin{figure}
\epsfxsize=7.0truecm\epsfbox{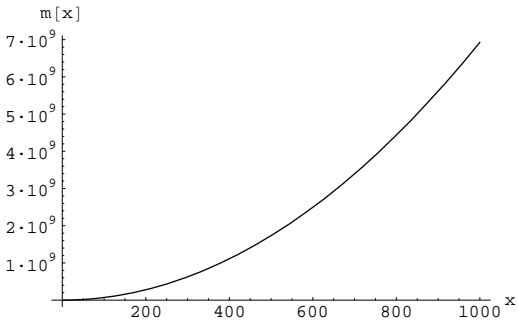} \narrowtext \caption{Mass
of the lens inside radius $x$ in the sinusoidal Swiss-cheese
model.}
\end{figure}

As expected, $m(x)$ increases as a function of $x$. Obviously
there is no solution to the equations $\lambda_r=0$ and
$\lambda_t=0$, as can be seen from the Fig 4 and Fig 5.
$\lambda_r=0$ may have a solution in the large '$x$' which is by
far out of the range of our approximation.

The result of no arcs, neither radial nor tangential, means that
the onion Swiss cheese model with sinusoidal solution is ruled out
by observations.


\begin{figure}
\epsfxsize=7.0truecm\epsfbox{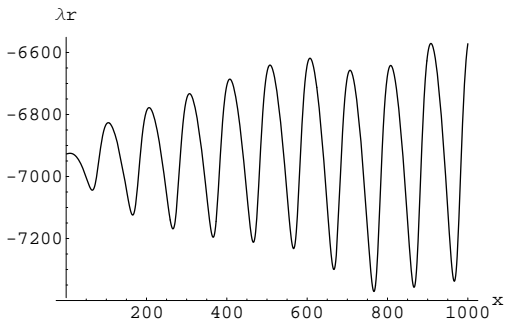} \narrowtext
\caption{Behavior of $\lambda_r(x)$ with respect to $x$ in the
Swiss-cheese model.}
\end{figure}


\begin{figure}
\epsfxsize=7.0truecm\epsfbox{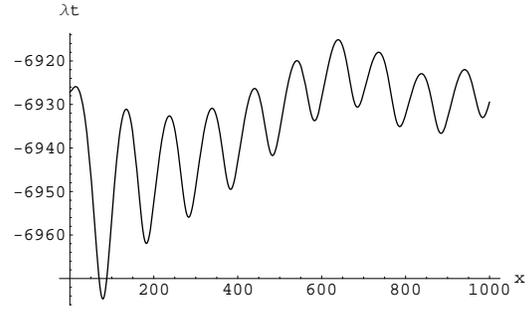} \narrowtext
\caption{Behavior of $\lambda_t(x)$ with respect to $x$ in the
Swiss-cheese model.}
\end{figure}

\subsection{Radial and tangential arcs II:  Swiss cheese model of Marra et
al}\label{kolb model}

We assume again that the time of light passage through the hole is
negligible relative to time evolution of the density of the hole in
the model of MKMR \cite{Kolb}. To derive the solution of Einstein
equations within the holes obeying the junction conditions, they
choose the initial density function to have a gaussian profile:

\ba \rho(r,t_i)&=&A\exp[-(r-r_M)^2/2\sigma^2]+\epsilon\:\:(r<r_h)\nonumber\\
\rho(r,t_i)&=&\rho_{FRW}(t_i)\:\:(r>r_h),\ea
where $\epsilon=0.0025$, $r_h=0.42$, $\sigma=r_h/10$, $r_M=.037$, $A=50.59$, and
$\rho_{FRW}(t_i)=25$. The hole ends at $r_h=.042$ which is equivalent to $350\:MPc$.
This is not a big hole but is almost an empty region: the matter density in the hole
is $10^4$ times smaller than in the cheese.

Applying this initial condition to the Einstein equations (in the
curved LTB case: $E(r)\neq0$), one gets $v(r,t)$ (the peculiar
velocity), $R(r,t)$, and $\rho(r,t)$. Hence, we have the density
profile of the lens at $t=0$ which, in their notation, is the
present time (Fig 6).

\begin{figure}
\epsfxsize=9.0truecm\epsfbox{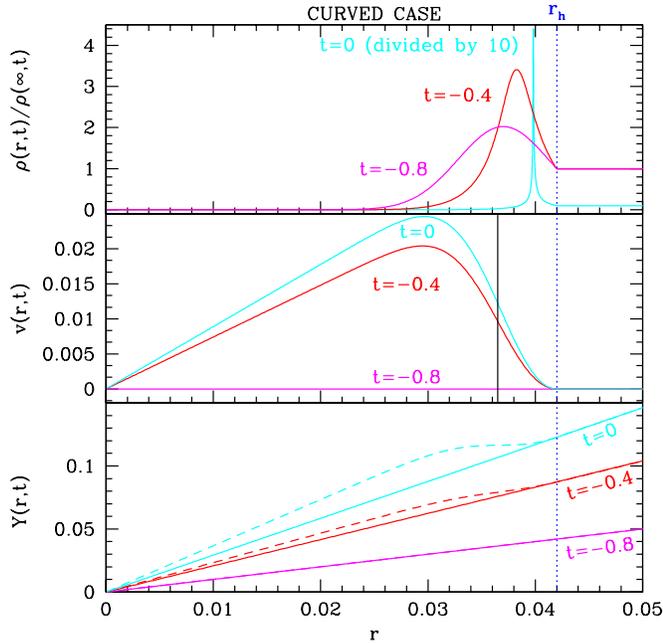} \narrowtext
\caption{Behavior of $R(r,t)(Y(r,t))$ in MKMR model, peculiar
velocity $v(r,t)$ and the density profile $\rho(r,t)$ with respect
to $r$ for the curved case at $t_i=-.8$ and $t=0$ (present time).
The straight lines in the $R(r,t)$ diagram are FRW solutions [4].}
\end{figure}

We have done the calculation along the same line as in the last
section, assuming the source and the observer far from the hole
and in the FRW background (cheese). As can be seen from the Figs 7
and 8, this model allows both radial and tangential arcs. However,
this happens at $r>r_h$ which means that the arcs will be observed
out of the inhomogeneous region.

\begin{figure}
\epsfxsize=7.0truecm\epsfbox{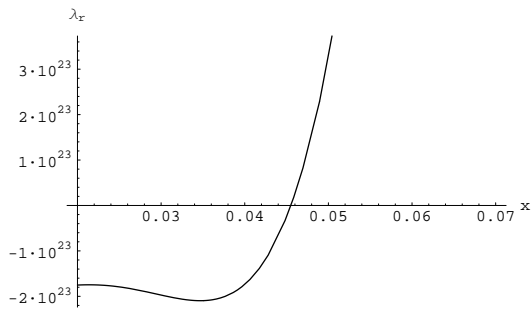} \narrowtext
\caption{Behavior of $\lambda_r(x)$ with respect to $x$ in Marra
et al model.}
\end{figure}


\begin{figure}
\epsfxsize=7.0truecm\epsfbox{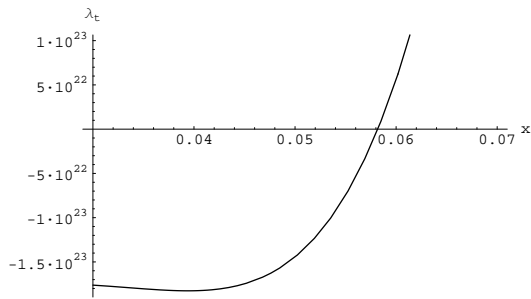} \narrowtext
\caption{Behavior of $\lambda_t(x)$ with respect to $x$ in Marra
et al model.}
\end{figure}

\section{Conclusion and Discussion}

Independent of how successful the inhomogeneous models are in
explaining the dark energy problem, gravitational lensing may serve
as a criterion to distinguish inhomogeneous cosmological models.
Different concepts developed in the cosmological gravitational
lensing techniques such as shear, convergence, tangential and radial
arcs, and time delays maybe used to see how tenable these models
are. The widely discussed LTB cosmological models, having a
vanishing shear as the FRW models, differ from FRW ones in the value
of convergence which may lead to observable effects such as
different time delays of the cosmological images and the large scale
lensing effects in the CMB. Assuming the observer outside the center
of symmetry of LTB, one expect a universal shear not
seen in the FRW models. \\
The Swiss cheese models provide us with a density profile for a
"hole", to be compared with the NFW profile. Therefore, the
question of tangential and radial arcs may lead us to a test of
such models, or to a better fixing of the model parameters. The
onion model predicts neither a tangential nor a radial arc. We may
therefore rule it out even as a toy model to explain dark energy.
The MKMR Swiss-cheese model \cite{Kolb} do produce both radial and
tangential arcs. The arcs are located in the cheese outside the
hole near the massive shell. The size of the hole is about 350
Mpc, much bigger than familiar structures in the universe.
Therefore, it is not possible to compare this result with real
data. It may be possible, however, to fix the parameter of the
model such that more realistic arcs results. If these parameters
are compatible with the explanation of the dimming of the
supernovas is another question. We therefore conclude that it is
desirable to do more research on different aspects of
gravitational lensing effects in inhomogeneous
models of the universe.\\

\section*{Acknowledgement}
SK-M thanks IPM astronomy school for hospitality. RM would like to
thank Iran TWAS chapter and ISMO for financial support.


\end{document}